\documentclass[12pt]{article}

\ifx\pdfoutput\undefined
\usepackage[dvips,bookmarks]{hyperref}
\else
\usepackage{hyperref}
\fi
\hypersetup{colorlinks=false,bookmarksopen,bookmarksnumbered,citecolor=blue,
   pdfstartview=FitH}

\usepackage{latexsym}
\usepackage{amssymb,amsfonts,amsmath}
\usepackage{graphicx} 
\usepackage{indentfirst}
\usepackage{bbm}
\usepackage{amssymb}
\usepackage{verbatim}
\usepackage{amsmath, amsthm,amssymb}
\usepackage{mathrsfs}
\usepackage{hyperref}
\usepackage{amsfonts}
\usepackage{dsfont}

\parskip=\medskipamount

\newcommand {\cA}{{\cal A}}

\newcommand {\cD}{{\cal D}}

\newcommand {\cF}{{\cal F}}
\newcommand {\cG}{{\cal G}}

\newcommand {\cK}{{\cal K}}

\newcommand {\cM}{{\cal M}}
\newcommand {\cN}{{\cal N}}

\newcommand {\cR}{{\cal R}}

\def\a{\alpha}
\def\b{\beta}

\def\d{\delta}

\def\f{\phi}
\def\g{\gamma}
\def\G{\Gamma}

\def\k{\kappa}

\def\m{\mu}

\def\o{\omega}

\def\q{\theta}

\def\x{\xi}

\def\D{\Delta}

\def\L{\Lambda}

\def\S{\Sigma}

\def\ri{{\rm i}}

\newcommand{\ve}{\varepsilon}                            

\newcommand{\pa}{\partial}                           
\newcommand{\hf}{\frac12}

\newcommand{\be}{\begin{equation}}
\newcommand{\ee}{\end{equation}}
\newcommand{\bea}{\begin{eqnarray}}
\newcommand{\eea}{\end{eqnarray}}
\newcommand{\non}{\nonumber}
\newcommand{\ba}{\begin{array}}
\newcommand{\ea}{\end{array}}

\newcommand{\dsR}{{\mathbb R}}

\newcommand{\bm}[1]{\mbox{\boldmath$#1$}}

\def\double #1{#1{\hbox{\kern-2pt $#1$}}}

\newcommand{\bsubeq}{\begin{subequations}}
\newcommand{\esubeq}{\end{subequations}}

\newcommand{\eps}{{\ve}}

\newcommand{\rd}{\mathrm d}
\newcommand{\de}{\nabla}

\numberwithin{equation}{section}

\renewcommand{\eps}{\ve}

\newcommand{\RM}{R(M)}
\newcommand{\RD}{R(\mathbb D)}
\newcommand{\RN}{R(N)}

\newcommand{\RS}{R(S)}
\newcommand{\RK}{R(K)}

\begin{document}
\begin{titlepage}
\begin{flushright}
August, 2013\\
\end{flushright}

\begin{center}
{\Large \bf 
$\bm{\cN = 6}$ superconformal gravity 
in three dimensions
from superspace
}
\end{center}

\begin{center}

{\bf
Sergei M. Kuzenko,
Joseph Novak and Gabriele Tartaglino-Mazzucchelli
} \\
\vspace{5mm}

\footnotesize{
{\it School of Physics M013, The University of Western Australia\\
35 Stirling Highway, Crawley W.A. 6009, Australia}}  
~\\
\texttt{joseph.novak,\,gabriele.tartaglino-mazzucchelli@uwa.edu.au}\\
\vspace{2mm}

\end{center}

\begin{abstract}
\baselineskip=14pt
A unique feature of $\cN=6$ conformal supergravity in  three dimensions
is that the super Cotton tensor $W^{IJKL}$ can equivalently be viewed, via the Hodge duality, 
as the field strength of an Abelian vector multiplet,  $W^{IJ}$. Using this observation and the conformal 
superspace techniques developed in arXiv:1305.3132 and arXiv:1306.1205, we construct the off-shell action 
for $\cN=6$ conformal supergravity. The complete component action is also worked out.
\end{abstract}

\vfill
\end{titlepage}

\newpage
\renewcommand{\thefootnote}{\arabic{footnote}}
\setcounter{footnote}{0}

\tableofcontents{}
\vspace{1cm}
\bigskip\hrule


\section{Introduction}

Recently, we have constructed the superspace actions for 
three-dimensional (3D) conformal supergravity theories with $\cN<6$ 
\cite{BKNT-M2}, and also worked out the complete component actions. 
Our construction made use of 
 the novel off-shell formulation for  $\cN$-extended 
conformal supergravity presented in \cite{BKNT-M1} and called
3D $\cN$-extended conformal superspace.\footnote{The 3D $\cN$-extended
conformal superspace \cite{BKNT-M1} is a generalization of the off-shell formulations
  for $\cN=1$ and $\cN=2$ conformal supergravity theories 
  in four dimensions \cite{ButterN=1,ButterN=2}.}
Upon degauging of certain local symmetries, conformal superspace 
reduces to the conventional formulation  
for $\cN$-extended conformal supergravity \cite{HIPT,KLT-M11}
with the structure group $\rm SL(2 , \dsR) \times SO(\cN)$.
However, the former formulation proves to be much more 
efficient for addressing certain problems such as 
the construction of conformal supergravity actions.

At the component level, the $\cN=1$ and $\cN=2$ supergravity actions given 
in \cite{BKNT-M2} coincide with those constructed in the 1980s
in \cite{vanN85} and  \cite{RvanN86}, respectively,
 using the superconformal tensor calculus.  But the off-shell $\cN=3$, $\cN=4$ and $\cN = 5$ supergravity actions 
were derived in \cite{BKNT-M2} for the first time. 
The method employed in \cite{BKNT-M2} broke down for $\cN>5$ due to a technical 
reason to be discussed below. In this note we make use of a unique property 
of the $\cN=6$ case to construct the corresponding off-shell action from superspace. 
Upon reduction to components, our action coincides with that found 
a week ago by Nishimura and Tanii
\cite{NT} who used completely different techniques.

This paper is organized as follows. 
In section 2  we briefly review $\cN=6 $ conformal supergravity in superspace 
and then describe the unique feature of $\cN=6 $ conformal supergravity.
In section 3 we make use of this property to construct the off-shell action for $\cN=6$ conformal supergravity, both 
in superspace and in terms of the component fields. 
The main results and their implications are discussed in section 4. 


\section{$\cN = 6$ conformal supergravity}

We begin with a brief review of $\cN=6 $ conformal superspace following 
 \cite{BKNT-M1}.  After that we describe the unique feature of $\cN = 6$ conformal supergravity 
 mentioned in the introduction. 

We consider
a curved three-dimensional $\cN = 6$  superspace
 $\cM^{3|12}$, parametrized by
local bosonic $(x^m)$ and fermionic coordinates $(\theta^\m_I)$, 
$ z^M = (x^m, \ \q^\mu_I) $,
where $m = 0, 1, 2$, $\mu = 1, 2$ and $I = 1, \cdots , 6$. 
The $\cN=6$ conformal superspace \cite{BKNT-M1}
 is obtained by gauging the $\cN=6$ superconformal algebra
${\mathfrak{osp}}(6|4, {\mathbb R})$ and then imposing appropriate constraints. 
The covariant derivatives have the form
\be
\nabla_A = E_A{}^M \pa_M - \o_A{}^{\underline b} X_{\underline b} 
= E_A{}^M \pa_M - \hf \Omega_A{}^{ab} M_{ab} - \hf \Phi_A{}^{PQ} N_{PQ} - B_A \mathbb D - \mathfrak{F}_A{}^B K_B \ .
\ee
Here $E_A = E_A{}^M(z) \partial_M$ is the inverse supervielbein, 
$M_{ab}$ are the Lorentz generators, $N_{IJ}$ are generators of the 
$\rm SO(6)$ group, $\mathbb D$ is the dilatation generator and $K_A = (K_a , S_\a^I)$ are the special superconformal 
generators.\footnote{As usual, we refer to $K_a$ as the special conformal generator and $S_\a^I$ as the $S$-supersymmetry generator.}
The complete set of the generators of 
${\mathfrak{osp}}(6|4, {\mathbb R})$
consists of  $X^{\tilde{a}} = ( P_A, X_{\underline{a}})$, where $P_A = (P_a, Q_\a^I)$ 
are the super-Poincar\'e generators. 

The Lorentz generators obey
\bsubeq
\bea
&[M_{ab} , M_{cd}] = 2 \eta_{c[a} M_{b] d} - 2 \eta_{d [a} M_{b] c} \ , \\
&[M_{ab} , \nabla_c ] = 2 \eta_{c [a} \nabla_{b]} \ , \quad [M_{\a\b} , \nabla_\g^I] = \eps_{\g(\a} \nabla_{\b)}^I \ .
\eea
\esubeq
The $\rm SO(6)$ and dilatation generators obey
\bsubeq
\begin{align}
[N_{KL} , N^{IJ}] &= 2 \d^I_{[K} N_{L]}{}^J - 2 \d^J_{[K} N_{L]}{}^I \ , \quad [N_{KL} , \nabla_\a^I] = 2 \d^I_{[K} \nabla_{\a L]} \ ,  \\
[\mathbb D , \nabla_a] &= \nabla_a \ , \quad [\mathbb D , \nabla_\a^I] = \hf\nabla_\a^I \ .
\end{align}
\esubeq
The special conformal generators $K_A$ transform under Lorentz and $\rm SO(6)$ transformations as
\bsubeq
\begin{align}
[M_{ab} , K_c] &= 2 \eta_{c[a} K_{b]} \ , \quad [M_{\a\b} , S_\g^I] = \eps_{\g(\a} S_{\b)}^I \ , \\
[N_{KL} , S_\a^I] &= 2 \d^I_{[K} S_{\a L]} \ ,
\end{align}
\esubeq
while under dilatations as
\begin{align}
[\mathbb D , K_a] = - K_a \ , \quad [\mathbb D, S_\a^I] &= - \hf S_\a^I \ .
\end{align}
Among themselves, the generators $K_A$ obey the algebra
\begin{align}
\{ S_\a^I , S_\b^J \} = 2 \ri \d^{IJ} (\g^c)_{\a\b} K_c \ .
\end{align}
Finally, the algebra of $K_A$ with $\nabla_A$ is given by
\bsubeq
\begin{align}
[K_a , \nabla_b] &= 2 \eta_{ab} \mathbb D + 2 M_{ab} \ , \\
[K_a , \nabla_\a^I ] &= - \ri (\g_a)_\a{}^\b S_\b^I \ , \\
[S_\a^I , \nabla_a] &= \ri (\g_a)_\a{}^\b \nabla_{\b}^I \ , \\
\{ S_\a^I , \nabla_\b^J \} &= 2 \eps_{\a\b} \d^{IJ} \mathbb D - 2 \d^{IJ} M_{\a \b} - 2 \eps_{\a \b} N^{IJ} \ .
\end{align}
\esubeq
The remaining (anti-)commutators are not essential here and may be found in \cite{BKNT-M1}.

Under the supergravity gauge group, the covariant derivatives transform as
\bea 
\d_\cG \nabla_A &=& [\cK , \nabla_A] \ ,
\label{TransCD}
\eea
where $\cK$ denotes the first-order differential operator
\bea
\cK = \xi^C \nabla_C + \hf \L^{ab} M_{ab} + \hf \L^{IJ} N_{IJ} + \L_{\mathbb D} \mathbb D + \L^A K_A \ .
\eea
Covariant (or tensor) superfields transform as
\bea 
\d_{\cG} T &=& \cK T~.
\eea

The covariant derivatives obey the (anti-)commutation relations of the form
\begin{align}
[ \nabla_A , \nabla_B \}
&= -T_{AB}{}^C \nabla_C - R_{AB}{}^{\underline{a}} X_{\underline{a}} \non\\
	&= -T_{AB}{}^C \nabla_C
	- \frac{1}{2} \RM_{AB}{}^{cd} M_{cd}
	- \frac{1}{2} \RN_{AB}{}^{PQ} N_{PQ}
	\non \\ & \quad
	- \RD_{AB} \mathbb D
	- \RS_{AB}{}^\g_K S_\g^K
	- \RK_{AB}{}^c K_c~,
\end{align}
where $T$ is the torsion, and $\RM$, $\RN$, $\RD$, $\RK$ are the curvatures.

The algebra of covariant derivatives corresponding to $\cN=6$ conformal supergravity is
\begin{subequations} \label{covDN>3}
\begin{align} 
\{ \nabla_\a^I , \nabla_\b^J \} &= 2 \ri \d^{IJ} \nabla_{\a\b} + \ri \eps_{\a\b} W^{IJKL} N_{KL} - \frac{\ri}{3} \eps_{\a\b} (\nabla^\g_K W^{IJKL}) S_{\g L} \non\\
&\qquad + \frac{1}{24} \eps_{\a\b} (\g^c)^{\g\d} (\nabla_{\g K} \nabla_{\d L} W^{IJKL}) K_c \ , \\
[\nabla_a , \nabla_\b^J ] &= \frac{1}{6} (\g_a)_{\b\g} (\nabla^\g_K W^{JPQK}) N_{PQ} \non\\
&\qquad - \frac{1}{24} (\g_a)_{\b\g} (\nabla^\g_L \nabla^\d_P W^{JKLP}) S_{\d K} \non\\
&\qquad - \frac{\ri}{240} (\g_a)_{\b\g} (\g^c)_{\d\rho} (\nabla^\g_K \nabla^\d_L \nabla^\rho_P W^{JKLP}) K_c \ , \\
[\nabla_a , \nabla_b] &
=    \frac{1}{48}   \eps_{abc} (\g^c)_{\a\b} 
\Big( \ri (\nabla^\a_I \nabla^\b_J W^{PQIJ}) N_{PQ} \non\\
&\qquad + \frac{\ri}{ 5} (\nabla^\a_I \nabla^\b_J \nabla^\g_K W^{LIJK}) S_{\g L} \non\\
&\qquad + \frac{1}{60} (\g^d)_{\g\d} (\nabla^\a_I \nabla^\b_J \nabla^\g_K \nabla^\d_L W^{IJKL}) K_d \Big) \ ,
\end{align}
\end{subequations}
where $W^{IJKL}= W^{[IJKL]} $ is the  super Cotton tensor, 
which is a  completely antisymmetric primary superfield of dimension 1, 
\be
S_\alpha^P W^{IJKL} = 0~, \quad \mathbb D W^{IJKL} = W^{IJKL} \ .
\ee
It satisfies the Bianchi identity
\be \nabla_{\a}^I W^{JKLP} 
= \nabla_\a^{[I} W^{JKLP]} - \frac{4}{3} \nabla_{\a Q} W^{Q [JKL} \d^{P] I} \ . 
\label{CSBIN>3}
\ee

The $\cN = 6$ case is special because it has the important property that the Hodge dual of the 
Cotton tensor
\be 
W^{IJ} := \frac{1}{4!} \eps^{IJKLPQ} W_{KLPQ}
\label{2.21}
\ee
satisfies the Bianchi identity for the field strength of an Abelian $\cN=6$ vector multiplet\footnote{The 
$\cN$-extended vector multiplet in conformal superspace is described in \cite{BKNT-M1}.}
\be 
\nabla_\a^{[I} W^{JK]} = \nabla_\a^{[I} W^{JK]} - \frac{2}{5} \d^{I [J} \nabla_{\a L} W^{K] L}~. \label{BIVM}
\ee
Therefore,
associated with the $\cN=6$ Weyl multiplet is a uniquely defined 
Abelian $\cN=6$ vector multiplet. 

As a result we can use  $W^{IJ}$ to define a closed two-form 
$F= \hf E^B \wedge E^A F_{AB }$, $\rd F =0$, 
with components
\bsubeq \label{FStwo-form}
\bea
F_{\a}^I{}_\b^J &=&  -2\ri\ve_{\a\b}W^{IJ} \ , \\
F_{a}{}_\a^I&=&
\frac{1}{ 5}(\g_a)_\a{}^{\b}\nabla_{\b J} W^{I J}
~,
\\
F_{ab}&=&
-\frac{\ri}{ 120}\ve_{abc}(\g^c)^{\a\b}[\nabla_{\a}^{ K},\nabla_{\b}^{ L}] W_{ K L}
~.
\eea
\esubeq
It is the field strength of the vector multiplet.  
We associate with the field strength  
$F$ a gauge one-form $A$,
\be F = \rd A \ .
\ee
It is the existence of this gauge one-form which distinguishes the $\cN = 6$ case from the $\cN < 6$ cases.


\section{Conformal supergravity action}

In this section, we start by recalling the method to construct the $\cN<6$ 
conformal supergravity actions employed in \cite{BKNT-M2}. 
After that, we  present a generalization of the method that is suitable in the 
$\cN=6$ case. 

\subsection{Construction of $\cN<6$ conformal supergravity actions}

The idea of the method employed in \cite{BKNT-M2}
is to look for two solutions, ${\bm \S}_{\rm CS}$  and  ${\bm \S}_R$, to the superform equation
\be \rd \bm \S = \langle R^2 \rangle \ ,
\ee
where\footnote{Following the notation of \cite{BKNT-M2}, we define $R^{\tilde{a}} := (R(P)^A , R^{\underline{a}})$ 
where $R(P)^{A} := T^A - \hat{T}^A$ and $\hat{T}^A$ is the flat superspace torsion.} 
\bea
 \langle R^2 \rangle := R^{\tilde{b}} \wedge R^{\tilde{a}} \G_{\tilde{a} \tilde{b}} \ , \quad \rd \langle R^2 \rangle = 0 
 \eea
 and $\G_{\tilde{a} \tilde{b}} $ denotes a properly normalized Cartan-Killing metric 
 of the $\cN$-extended superconformal algebra;
the explicit form of $ \langle R^2 \rangle $ is
\be
 \langle R^2 \rangle 
=  - R(N)^{IJ} \wedge R(N)_{IJ} \label{SE1}~.
\ee
One solution is always the Chern-Simons form $\bm \S_{\rm CS}$ 
which exists for any $\cN$. 
Its explicit form is found to be
\begin{align} {\bm \S}_{\rm CS} &=
	- {\hat{R}}^a \wedge \Omega_a  -  \frac{1}{6} \Omega^c \wedge \Omega^b \wedge \Omega^a \eps_{abc}
	- 4 \ri E^a \wedge \frak{F}^{\a I} \wedge \frak{F}^\b_I (\g_a)_{\a\b}
	- {\hat{R}}^{IJ} \wedge \Phi_{IJ}
	\non\\ & \quad
	+ \frac{1}{3} \Phi^{IJ} \wedge \Phi_I{}^K \wedge \Phi_{KJ}
	+ 2 E^a \wedge \frak{F}_a \wedge B
	- 2 E^{\alpha}_I \wedge \frak{F}_\alpha^I \wedge B
	+ {\rm exact \ form} \ , \label{CSFORM}
\end{align}
where
\begin{align}
{\hat{R}}^{ab} := \rd \Omega^{ab} + \Omega^{ac} \wedge \Omega_c{}^b ~, \qquad
{\hat{R}}^{IJ} := \rd \Phi^{IJ} + \Phi^{IK} \wedge \Phi_{K}{}^J 
\end{align}
correspond to the Riemann and  $\rm SO(\cN)$ curvature tensors.
The other solution
is the so-called curvature induced form $\bm \S_R$ such that its components are
constructed in terms of the super Cotton tensor
and its covariant derivatives. It turns out that  $\bm \S_R$ exists for $\cN<6$
(in the cases $\cN=1$ and $\cN=2$, $\bm \S_R$ vanishes), see 
\cite{BKNT-M2} for more details.
 
The difference 
\be
 \frak{J} = \bm \S_{\rm CS} - \bm \S_{R} 
\ee
is a closed three-form which may be used to construct a locally supersymmetric action
\bea 
S = \int_{\cM^3} \frak{J} = \int \rd^3 x \,e \,{}^* \frak{J} |_{\q=0}\ , \qquad 
{}^*\frak{J} = \frac{1}{3!} \eps^{mnp} \frak{J}_{mnp} ~.
\label{ectoS}
\eea
This action principle is not applicable for $\cN \geq 6$ since $\bm \S_R$ does not exist.
However, in the $\cN=6$ case 
there is a way out and that is to make use of the
closed  two-form \eqref{FStwo-form}
constructed in terms of the vector multiplet field strength  $W^{IJ}$, 
eq. \eqref{2.21}. 

\subsection{Modified $\cN=6$ curvature induced form}

In the $\cN=6$ case, we can modify the superform equation \eqref{SE1} to
\be 
\rd \S = - R(N)^{IJ} \wedge R(N)_{IJ} - \cA F \wedge F \ , \label{SE2}
\ee
where $\cA$ is some constant we will determine. Here $F$ is 
the closed two-form \eqref{FStwo-form}.
The Chern-Simons solution is now modified to
\begin{align} 
\S_{\rm CS} &=
	- {\hat{R}}^a \wedge \Omega_a  -  \frac{1}{6} \Omega^c \wedge \Omega^b \wedge \Omega^a \eps_{abc}
	- 4 \ri E^a \wedge \frak{F}^{\a I} \wedge \frak{F}^\b_I (\g_a)_{\a\b}
	- {\hat{R}}^{IJ} \wedge \Phi_{IJ}
	\non\\ & \quad
	+ \frac{1}{3} \Phi^{IJ} \wedge \Phi_I{}^K \wedge \Phi_{KJ}
	+ 2 E^a \wedge \frak{F}_a \wedge B
	- 2 E^{\alpha}_I \wedge \frak{F}_\alpha^I \wedge B - \cA F \wedge A 
	\non\\ & \quad
	+ {\rm exact \ form} \ .
 \label{CSFORM2}
\end{align}

We may now attempt to find a covariant solution to eq. \eqref{SE2} 
which will also be called the curvature induced form and denoted $\S_R$.  
We make the ansatz for the 
lowest components\footnote{When referring to components of the curvature induced form we 
denote $\S_R$ by $\S$ to avoid awkward notation.}
\be \S_\a^I{}_\b^J{}_\g^K = 0 \ , \quad \S_a{}_\b^J{}_\g^K = \ri (\g_a)_{\b\g} (A \d^{JK} W^{PQ} W_{PQ} + B W^{JP} W^K{}_P) \ ,
\ee
with $A$ and $B$ some constants to be determined,  
and turn to analyzing the superform equation \eqref{SE2} by increasing dimension.

At the lowest dimension we find that we must set $\cA = - 2$ and
\bsubeq
\be \S_a{}_\b^J{}_\g^K = 8 \ri (\g_a)_{\b\g} (W^{JP} W^K{}_P - \frac{1}{4} \d^{JK} W^{PQ} W_{PQ}) \ .
\ee
The higher dimension components are found to be
\begin{align} 
\S_{ab}{}_\g^K &= - 2 \eps_{abc} (\g^c)_\g{}^\d \Big( \nabla_\d^{[K} W^{PQ]} W_{PQ} - \frac{2}{5} \nabla_{\d}^P W_{Q P} W^{Q K} \Big) \ , \\
\S_{abc} &= \ri \eps_{abc} \Big( \frac{1}{5} \nabla^{\g I} \nabla_{\g K} W^{JK} W_{IJ} + \frac{1}{3} \nabla^{\g [I} W^{JK]} \nabla_{\g [I} W_{JK]} 
- \frac{2}{25} \nabla^\g_I W^{KI} \nabla_\g^J W_{KJ} \non\\
&\qquad \qquad - \frac{\ri}{3} \eps^{IJKLPQ} W_{IJ} W_{KL} W_{PQ} \Big) \ .
\end{align}
\esubeq
To derive these results,
we have made use of the following consequences of the Bianchi identity \eqref{BIVM}:
\bsubeq
\begin{align}
\nabla_\a^I \nabla_{\b K} W^{JK} &= \hf \eps_{\a\b} \nabla^{\g [I} \nabla_{\g K} W^{J] K} - 5 \ri \nabla_{\a\b} W^{IJ} 
+ \frac{1}{6} \d^{IJ} \nabla_{(\a K} \nabla_{\b ) L } W^{KL} \ , \\
\nabla^\g_P \nabla_\g^P W^{IJ} &= \frac{2}{5} \nabla^{\g [I} \nabla_{\g K} W^{J] K} - 4 \ri W^{IJKL} W_{KL} \ , \\
\nabla_\a^I \nabla_\b^{[J} W^{KL]} &= \nabla_{(\a}^{[I} \nabla_{\b)}^J W^{KL]} + \frac{3}{10} \eps_{\a\b} \d^{I [J} \nabla^{\g K} \nabla_{\g P} W^{L] P} 
\non\\
&\quad + 3 \ri \d^{I [J} \nabla_{\a\b} W^{KL]} + 6 \ri \eps_{\a\b} W^{PI [JK} W^{L]}{}_P \ .
\end{align}
\esubeq
Now, the closed three-form $ \frak{J} =  \S_{\rm CS} - \S_{R} $ generates
a locally supersymmetric action according to the rule \eqref{ectoS}.


\subsection{The component action}

The complete component analysis of the $\cN$-extended Weyl multiplet
was given in \cite{BKNT-M2}. Here we specialize to the $\cN = 6$ case where the auxiliary fields 
coming from the super Cotton tensor are defined as:
\begin{subequations} \label{defCompW}
\begin{align}
w_{IJKL} &:= W_{IJKL}|
\equiv\frac{1}{2}\ve_{IJKLPQ}w^{PQ} \ , \\
w_\a{}^{IJK} &:= - \frac{\ri}{6} \nabla_{\a L} W^{IJKL}|
\equiv\frac{1}{3!}\ve^{IJKLPQ}\tilde{w}_\a{}_{LPQ} \ , \\
y^{I J K L} &:= \frac{\ri}{3} \nabla^{\g [I}  \nabla_{\g P} W^{JKL]P}| 
\equiv\frac{1}{2}\ve^{IJKLPQ}y_{PQ}\ , \\
X_{\a}{}^{I_1 \cdots I_{5}} &:= \ri \nabla_{\a}^{[I_1}  W^{I_2\cdots I_{5}]} 
\equiv\ve^{I_1\cdots I_5J}X_\a{}_{J}
\ . 
\label{3.13d}
\end{align}
\end{subequations}
These definitions agree with  \cite{GreitzHowe,GGHN}. There is also an additional component field 
$X_{\a\b}{}^{I_1 \cdots I_6}:= \ri \nabla_{(\a}^{[I_1} \nabla_{\b)}^{I_2} W^{I_3\cdots I_{6}] }
 = - \eps^{I_1 \cdots I_6} F_{\a\b}|$. 
However, this field turns out to be a composite object as it is 
the component U(1) field strength $\cF_{ab}$
up to contributions involving the gravitino:
\bea
F_{ab}|
=
\cF_{ab}
+\ri \psi_{[a}{}^\b_J (\g_{b]})_\b{}^\g X_\g{}^J
-\hf\ri \psi_{[a}{}^\a_I \psi_{b]}{}_{\a J}w^{IJ}
~,
~~~
\cF_{ab}:=
2e_{a}{}^me_b{}^n\pa_{[m}A_{n]}
~.~~~~~~
\eea

As the action is invariant with respect to the gauge transformations \eqref{TransCD} 
up to a total derivative, it follows that the dependence
on $b_m$ must drop out. Equivalently, we can simply
adopt the $K$-gauge $b_m=0$. 
Using the action \eqref{ectoS} and the
Chern-Simons form \eqref{CSFORM2}, we find the Chern-Simons contribution to be
\begin{align}\label{eq:Scs}
S_{\rm CS} =& \ \frac{1}{4} \int \rd^3 x \,e \,\eps^{abc} \,\Big( \omega_{a}{}^{fg} \cR_{bc}{}_{fg}  - \frac{2}{3} \omega_{af}{}^g \omega_{bg}{}^h \omega_{ch}{}^f 
-  \frac{\ri}{2} \Psi_{bc}{}^\a_I (\g_d)_\a{}^\b (\g_a)_\b{}^\g \eps^{def} \Psi_{ef}{}^I_\g  \non\\
&- 2 {\cR}_{ab}{}^{IJ} V_c{}_{IJ} - \frac{4}{3} V_a{}^{IJ} V_b{}_I{}^K V_c{}_{KJ} + 4 \cF_{ab} A_c \Big) \ ,
\end{align}
where the component curvatures $\cR_{ab}{}^{cd}$ and $\cR_{ab}{}^{IJ}$  are defined as 
\bsubeq
\begin{align}
\cR_{ab}{}^{cd} &:= 2 e_a{}^m e_b{}^n \partial_{[m} \omega_{n]}{}^{ab} - 2 \omega_{[a}{}^{cf} \omega_{b]}{}_f{}^d 
~,
\\
{\cR}_{ab}{}^{IJ} &:= 2 e_a{}^m e_b{}^n \partial_{[m} V_{n]}{}^{IJ} - 2 V_{[a}{}{}^{IK} V_{b] K}{}^J \ .
\end{align}
\esubeq

Using the formula
\begin{align}\label{eq:SigmaProjection}
\frac{1}{3 !} \eps^{mnp} {\S}_{mnp}| &=
	\frac{1}{3 !} \eps^{mnp} E_p{}^C E_n{}^B E_m{}^A {\S}_{ABC}| \non \\
	&= \frac{1}{3 !} \eps^{abc} \big( {\S}_{abc}| + \frac{3}{2} \psi_a{}^\a_I {\S}_\a^I{}_{bc}| + \frac{3}{4} \psi_b{}^\b_J \psi_a{}^\a_I {\S}_\a^I{}_\b^J{}_{c}|
	\non\\&\quad
	+ \frac{1}{8} \psi_c{}^\g_K \psi_b{}^\b_J \psi_a{}^\a_I {\S}_\a^I{}_\b^J{}_\g^K| \big)
\end{align}
for the component projection of a three-form along with the explicit
expressions for the components of $\S_{ABC}$, we find
\begin{align}
\frac{1}{3 !} \eps^{mnp} {\S}_{mnp}| &= y^{IJ} w_{IJ} + \frac{4 \ri}{3} \tilde{w}^\a{}^{IJK} \tilde{w}_\a{}_{IJK} - 2 \ri X^\g_K X_\g^K 
+ \frac{2}{3} \eps^{IJKLPQ} w_{IJ} w_{KL} w_{PQ} \non\\
&\quad+ 2 \ri \psi_a{}^\a_I (\g^a)_\a{}^\b (\tilde{w}_\b{}^{IJK} w_{JK} + X_{\b J} w^{IJ})\non\\
&\quad
 - \ri \eps^{abc} (\g_a)_{\a\b} \psi_b{}^\a_I \psi_c{}^\b_J (w^{IK} w^J{}_K - \frac{1}{4} \d^{IJ} w^{KL} w_{KL}) 
 \ ,
\end{align}
where we have used the relations
\begin{subequations} \label{defCompW2}
\begin{align}
\tilde{w}_\a{}^{IJK} &
= - \frac{\ri}{2} \nabla_\a^{[I} W^{JK]}| \ , \\
y^{I J} &
= - \frac{\ri}{5} \nabla^{\g [I}  \nabla_{\g P} W^{J]P}| - \frac{1}{2} \eps^{IJKLPQ} W_{KL} W_{PQ}| \ , \\
X_{\a}{}^{I} &
= - \frac{\ri}{5} \nabla_{\a J} W^{IJ}|\ .
\end{align}\end{subequations}

Combining this result with the Chern-Simons contribution gives the full action
\begin{align}
S =& \  \frac{1}{4} \int \rd^3 x \,e \,\Big\{ \eps^{abc} \big( \omega_{a}{}^{fg} \cR_{bc}{}_{fg} - \frac{2}{3} \omega_{af}{}^g \omega_{bg}{}^h \omega_{ch}{}^f 
-  \frac{\ri}{2} \Psi_{bc}{}^\a_I (\g_d)_\a{}^\b (\g_a)_\b{}^\g \eps^{def} \Psi_{ef}{}^I_\g 
	\non\\&\quad
	- 2 {\cR}_{ab}{}^{IJ} V_c{}_{IJ} - \frac{4}{3} V_a{}^{IJ} V_b{}_I{}^K V_c{}_{KJ} + 4 \cF_{ab} A_c \big)
	\non\\&\quad
	- 4 y^{IJ} w_{IJ} - \frac{16 \ri}{3} \tilde{w}^\a{}^{IJK} \tilde{w}_\a{}_{IJK} + 8 \ri X^\g_K X_\g^K - \frac{8}{3} \eps^{IJKLPQ} w_{IJ} w_{KL} w_{PQ} \non\\
&\quad- 8 \ri \psi_a{}^\a_I (\g^a)_\a{}^\b (\tilde{w}_\b{}^{IJK} w_{JK} + X_{\b J} w^{IJ})\non\\
&\quad + 4 \ri \eps^{abc} (\g_a)_{\a\b} \psi_b{}^\a_I \psi_c{}^\b_J (w^{IK} w^J{}_K - \frac{1}{4} \d^{IJ} w^{KL} w_{KL})\Big\} \ .
\label{3.20}
\end{align}

Our choice of normalization for the auxiliary fields allows a simple truncation
to lower values of $\cN$. From the above action one can truncate the
auxiliary fields to $\cN=5$ by taking (with $ I,J,K=1,2,3,4, 5 $)
\begin{align}\label{eq:5to4Trunc}
\left.\begin{gathered}
w_{IJ} \longrightarrow 0~, \qquad 
\tilde{w}_\alpha{}^{IJK} \longrightarrow 0~, \qquad
X_\alpha{}^I \longrightarrow 0~, \qquad
y^{IJ} \longrightarrow 0~, \\
w^{I6} \longrightarrow w^I~, \qquad 
\tilde{w}_\alpha{}^{IJ6} \longrightarrow w_\alpha{}^{IJ} ~, \qquad
y^{I6} \longrightarrow y^I ~
\end{gathered} \quad \right\} 
 \ .
\end{align}
For the gauge fields one must switch off the $\rm U(1)$ gauge field $A_b \longrightarrow 0$, while truncation is obvious for the other gauge fields. 
The $\cN < 5$ cases can be obtained via the truncation procedure given in \cite{BKNT-M2}.


\section{Discussion}

Our component action for $\cN=6 $ conformal supergravity \eqref{3.20} 
agrees with that derived recently in \cite{NT}, 
where alternative techniques were used involving the 
consistent truncation of the off-shell multiplet of $\cN=8$ conformal supergravity \cite{NT12}.  
It also correctly reduces to the action for $\cN=5$ conformal supergravity \cite{BKNT-M2}
via the truncation procedure \eqref{eq:5to4Trunc}. 
Eliminating the auxiliary fields is equivalent to removing the last three lines in 
\eqref{3.20}.  The resulting on-shell action for $\cN=6 $ conformal supergravity
does not agree with that obtained in \cite{LR89,NG} 
by gauging the $\cN=6$ superconformal algebra in $x$-space (the action given in \cite{LR89, NG} does not 
contain the $\rm U(1)$ Chern-Simons term). 
Instead it coincides with the action given in \cite{CN}.

In conclusion, we comment on the structure of a supercurrent multiplet 
associated with  a superconformal matter theory coupled to $\cN$-extended conformal supergravity
(see also \cite{BKNT-M1}).
In general, the supergravity-matter system is described by an action of the form
\bea
S= \frac{1}{\k} S_{\rm CSG} + S_{\rm matter}~,
\eea
where $S_{\rm CSG}$ denotes the conformal supergravity action and $S_{\rm matter}$ the matter action.
The conformal supergravity equation is 
\bea
\frac{1}{\k} W + T = 0~.
\eea
Here $W$ is the $\cN$-extended super Cotton tensor (with its indices suppressed)
 and $T$ the matter supercurrent multiplet. The supercurrent has the same algebraic type 
 as $W$ and obeys the same differential constraints $W$ is subject to. 
 For any $\cN$, the super Cotton tensor (and also the supercurrent) is a conformal primary superfield, 
 \bea
 S^I_\a W =0~, \qquad {\mathbb D} W = \D_W W~,
 \eea
 with $\D_W$ the dimension of $W$. We now recall the structure of $W$ for various values of $\cN$
 following \cite{BKNT-M1}. 
 
 The $\cN=1$ super Cotton tensor \cite{KT-M12}
 is a completely symmetric spinor
 $W_{\a\b\g}$ of dimension 5/2. It obeys the conformally invariant constraint
\be
\nabla^\a W_{\a \b\g} = 0 \ . \label{divLessN1}
\ee
In the $\cN=2$ case,  the super Cotton tensor \cite{ZP,Kuzenko12} 
is a completely symmetric spinor
 $W_{\a\b}$ of dimension 2. The corresponding conformally invariant constraint is 
\be
\nabla^{\a I} W_{\a\b} = 0 \ . 
\label{divLessN2}
\ee
In the $\cN=3$ case,  the super Cotton tensor is a symmetric spinor
 $W_{\a}$ of dimension 3/2. It obeys the conformally invariant constraint 
 \be
\nabla^{\a I} W_\a = 0 \ . \label{divLessN3}
\ee
For $\cN>5$,  the super Cotton tensor \cite{HIPT,KLT-M11}
is a completely antisymmetric tensor $W^{IJKL}$
of dimension 1. It obeys  the conformally invariant constraint
\be \nabla_{\a}^I W^{JKLP} = \nabla_\a^{[I} W^{JKLP]} - \frac{4}{\cN - 3} \nabla_{\a Q} W^{Q [JKL} \d^{P] I} \ .
\ee
In the $\cN=4$ case, the super Cotton tensor is equivalently described by a scalar primary 
dimension-1 superfield  $W^{IJKL}:=\ve^{IJKL}W$. 
The corresponding conformally invariant constraint is 
\bea
\nabla^{\a I}\nabla_{\a}^JW=\frac{1}{4}\d^{IJ}\nabla^{\a}_P\nabla_{\a}^PW~.
\eea
As we have demonstrated, the specific feature of the $\cN=6$ case is that the super Cotton
tensor is equivalent to the U(1)  vector multiplet field strength \eqref{2.21}. 
Therefore, the $\cN=6$ supercurrent $T^{IJ}$ has the same multiplet structure. 
This agrees with the Nishimura-Tanii analysis  \cite{NT} of the supercurrent of the 
ABJM model \cite{Aharony:2008ug} coupled to conformal supergravity.  
\\


\noindent
{\bf Acknowledgements:}\\
We are grateful to Daniel Butter for collaboration at the early stage of this project. 
The work of SMK and JN was supported in part by the Australian Research Council,
project No. DP1096372.  
The work of GT-M and JN was  supported in part by the Australian Research Council's Discovery Early Career 
Award (DECRA), project No. DE120101498.


\appendix

\section{The supersymmetry transformations} \label{SUSY}

{\allowdisplaybreaks

In this appendix we present the complete $Q$- and $S$-supersymmetry transformations for
the component fields of the Weyl multiplet for $\cN=6$.
The component action \eqref{3.20} is manifestly supersymmetric by virtue of our superspace construction.
We refer the reader to \cite{BKNT-M2} for details on the component projection rules
and the precise definition of the gauge component fields. 

The $Q$-supersymmetry transformations of the connections are
\begin{subequations}
\bea
\delta_Q(\xi) e_m{}^a &=& \ri \psi_m{}^{\a I} (\g^a)_\a{}^\b \xi_{I\b} \ , \\
\delta_Q(\xi) \psi_m{}^{\alpha I} &=&  2\cD_m \xi^{\alpha I} =
	2 \pa_m \xi^{\a I} + \omega_m{}^\a{}_\b \xi^{\b I}
	+ b_m \xi^{\a I}
	- 2 V_m{}^I{}_J \xi^{\a J}, \\
\delta_Q(\xi) b_m &=&
	-\phi_m{}^{\a J} \xi_{\a J} \ , \\
\delta_Q(\xi) V_m{}^{IJ} &=& 2 \phi_m{}^{\a[I} \xi_\a^{J]} 
- \frac{\ri}{2}\ve^{IJKLPQ} \psi_m{}^{\a}_{K} \xi_{\a L} w_{PQ}
	- \frac{\ri}{3} \ve^{IJKLPQ}\xi_K^\a (\gamma_m)_\a{}^\b \tilde{w}_\b{}_{LPQ} \ , 
~~~~~~~~~
\label{eq:dQphiN} \\
\delta_Q(\xi) \omega_m{}^{ab} &=&
	- \ve^{abc} \phi_m{}^{\a I} (\gamma_c)_\a{}^\b \xi_{\b I} \ , \\
\delta_Q(\xi) \phi_m{}^{\alpha I} &=&
	-2 \ri \xi^{\b I} (\gamma_b)_\b{}^\a \frak{f}_m{}^b
	+ \frac{1}{3} \ve^{IJKLPQ}\psi_{m}{}^{\b}_{ J} \xi_{\b K} \tilde{w}^{\a}{}_{ LPQ}
\non\\
&&
+ \ri \xi^{\b J} (\g_m)_\b{}^\g (\g_c)_\g{}^\a R(N)^{c J I}| 
\label{111111}
\ , 
\\
\delta_Q(\xi) \frak{f}_m{}^a &=& \frac{\ri}{2} \psi_m{}^{\a}_{I} \xi_{\a J} R(N)^{a IJ}|
	+ \xi_I^\a (\gamma_m)_\a{}^\b R(S)^{a}{}_\b^I| \ ,
	\label{111112}
	\\
\delta_{Q}(\x) A_m
&=&
-\ri \xi^\a_K (\g_m)_\a{}^\b X_\b{}^K
-\ri \psi_m{}^\b_J \xi_{\b K}w^{JK}
~.
\eea
\end{subequations}
Here we have made use of the covariant derivative
\be
\cD_a 
= 
e_a{}^m \cD_m = e_a{}^m (\partial_m - \hf \omega_m{}^{bc} M_{bc} - \hf V_m{}^{IJ} N_{IJ} - b_m \mathbb D)
\ee
and the following results for the projection of the SO(6) and $S$-supersymmetry 
curvatures
\bsubeq
\bea
R(N)_{ab}{}^{IJ}|
&=&\cR_{ab}{}^{IJ} 
+ 2 \psi_{[a}{}^{\a [I} \phi_{b]}{}_\a^{J]}
+ \frac{\ri}{3} \,\ve^{IJKLPQ}\psi_{[a}{}^{\beta}_K (\gamma_{b]})_\beta{}^\gamma \tilde{w}_{\gamma}{}_{LPQ}
\non\\
&&
- \frac{\ri}{4} \ve^{IJKLPQ}\psi_{[a}{}^{\beta}_K \psi_{b] \beta L} w_{PQ}
~,~~~~~~~~~
\\
R(S)^c{}^\a_I|
&=&
 e_{[a}{}^m e_{b]}{}^n \cD_m \phi_n{}^\a_I + \ri \,\psi_{[a}{}^\b_I \mathfrak f_{b]}{}^c (\g_c)_\b{}^\a
- \frac{\ri}{2} \psi_{[a}{}^{\b J}(\gamma_{b]})_\b{}^\g(\gamma_c)_\g{}^\alpha R(N)^{c}{}_{JI}| 
\non\\
&&
+ \frac{1}{12}\ve_{IJKLPQ} \psi_{[a}{}^{\b J} \psi_{b]}{}_{\b}^{K} \tilde{w}^{\alpha LPQ}
~.
\label{111113}
\eea
\esubeq
The curvature tensors $R(N)^c{}^{IJ}$ and $R(S)^c{}_\a^I$ 
in eqs. \eqref{111111}, \eqref{111112} and \eqref{111113}
are the Hodge-duals of 
$R(N)_{ab}{}^{IJ}$ and $R(S)_{ab}{}_\a^I$, respectively.\footnote{Given a two-form $F_{ab}$, 
its  Hodge-dual is $F^c = \frac{1}{2} \eps^{cab} F_{ab}$.} 
The $S$-supersymmetry transformations of the connections are
\begin{subequations}
\bea
\delta_S(\eta) e_m{}^a &=&
 0~, \\
\delta_S(\eta) \psi_m{}^{\alpha I} 
&=&
- 2\ri \,\eta^{\b I} (\gamma_m)_\b{}^\a \ , \\
\delta_S(\eta) b_m &=&
 \psi_m{}^{\a I} \eta_{\a I} \ , 
 \\
\delta_S(\eta) V_m{}^{IJ} &=&
 2 \psi_m{}^{\a[I} \eta_\a^{J]} \ , 
 \\
\delta_S(\eta) \omega_m{}^{ab} &= &
-\ve^{abc} \psi_m{}^{\a I}(\gamma_c)_\a{}^\b \eta_{\b I} \ , 
\\
\delta_S(\eta) \phi_m{}^{\alpha I} &=&
	2 \cD_m \eta^{\alpha I} =
	2 \pa_m \eta^{\alpha I} + \omega_m{}^{\alpha}{}_\beta \eta^{\beta I}
	- b_m \eta^{\alpha I}
	- 2 V_m{}^{I}{}_J \eta^{\alpha J} \ ,
	 \\
\delta_S(\eta) \frak{f}_m{}^a &= &\ri \,\phi_m{}^{\a I} (\gamma^a)_\a{}^\b \eta_{\b I} \ ,
\\
\delta_{S}(\eta) A_m
&=&0~.
\eea
\end{subequations}
It should be pointed out that all the auxiliary fields defined in \eqref{defCompW}, 
as well as the Cottino $w_{\a\b\g}{}^I$ and Cotton $w_{\a\b\g\d}$ tensors \cite{BKNT-M2},
are ordinary  primary fields annihilated by the special conformal generator $K_a$.

Along with the transformation laws of the gauge fields of the Weyl multiplet it is necessary to have the 
transformation rules of the auxiliary fields given by eqs. \eqref{defCompW} and \eqref{defCompW2}.
Their $Q$- and $S$-supersymmetry transformations are
\bsubeq
\bea
\d_Q(\x) w^{IJ}
&=&
2\ri\x^\a_K\tilde{w}_\a{}^{KIJ}\
-2\ri\x^{\a [I}X_\a{}^{J]}
~,
\\
\d_S(\eta) w^{IJ}&=&0
~,
\\
\d_Q\tilde{w}_\a{}^{IJK} &=&
\frac{1}{4} (\g^a)_{\a\b}\x^\b_P \ve^{PIJKST}R(N)_{a}{}_{ST}|
+\frac{3}{2} (\g^a)_{\a\b}\x^{\b [I}\hat{\de}_a w^{JK]}
-\frac{3}{4}\x_{\a}^{[I}y^{JK]}
~,~~~~~~
\\
\d_S\tilde{w}_\a{}^{IJK} 
&=&
 3\ri \eta_{\a}^{ [I}w^{JK]}
 ~,
 \\
 \d_QX_{\a}{}^{I} 
&=&
 \x^{\g I}F_{\g\a}|
-\frac{1}{2}\x_{\b K}y^{KI}
-\x^\g_K\hat{\de}_{\g\b}w^{KI}
+\frac{1}{4}\ve^{IJKLPQ}\x_{\a J}w_{KL}w_{PQ}
~,
\\
\d_SX_{\a}{}^{I} 
&=&
2\ri\eta_{\a K}w^{IK}
~,
\\
\d_Q y^{IJ}
&=&
\non
-4\ri\x^{\g[I}\hat{\de}_{\g\b}X^{\b J]}
+4\ri\x^\g_ K\hat{\de}_{\g}{}^{\b}\tilde{w}_\b{}^{KIJ}
+\frac{4\ri}{3}\x^\g_ K \tilde{w}_\g{}^{MNP}\ve_{MNP}{}^{K[I}{}_{Q}w^{J]Q}
\non\\
&&
-\frac{4\ri}{3}\tilde{w}_\g{}_{MNP}\,w_{QS}\,\ve^{MNPQS[I}\x^{\g J]}
~,
\\
\d_Sy^{IJ}
&=&
-4\eta^{\d [I}X_\d^{J]}
-4\eta^\d_L\tilde{w}_\d{}^{LIJ}
~,
\eea
\esubeq
where we have defined
\bsubeq
\bea
\hat{\de}_a w^{IJ}
&:=&
\cD_a w^{IJ}
-\ri\psi_a{}^\g_K\tilde{w}_\g{}^{KIJ}
-\ri\psi_a{}^{\g [I}X_\g{}^{J]}
~,
\\
\hat{\de}_{a}X^{\b J}
&:=&
\cD_aX^{\b J}
-\hf\psi_a{}_\g^{J}F^{\b\g}|
+\frac{1}{4}\psi_a{}^\b_Ly^{LJ}
+\hf(\g^b)^{\b\g}\psi_a{}_{\g L}\hat{\de}_b w^{LJ}
\non\\
&&
+\frac{1}{8}(\g^a)_{\a\b}\ve^{JKLPQS}\psi_a{}^\b_{K} w_{LP}w_{QS}
+\ri(\g^a)_{\a\b}\f_a{}^\b_{L}w^{JL}
~,
\\
\hat{\de}_a\tilde{w}_\b{}^{IJK}
&:=&
\cD_a\tilde{w}_\b{}^{IJK}
-\frac{1}{8}(\g^b)_{\b}{}^\g\ve^{IJKLPQ}\psi_a{}_{\g L} R(N)_{b}{}_{PQ}|
+\frac{3}{8}\psi_a{}_\b^{[I}y^{JK]}
\non\\
&&
+\frac{3}{4}(\g^b)_{\b}{}^{\d}\psi_a{}_{\d L} \hat{\de}_b w^{[IJ}\d^{K]L}
-\frac{3\ri}{2}\f_a{}_{\b}^{[I}w^{JK]}
~.
\eea
\esubeq

}


\begin{footnotesize}

\end{footnotesize}


\begin{thebibliography}{66}

\bibitem{BKNT-M2} 
  D.~Butter, S.~M.~Kuzenko, J.~Novak and G.~Tartaglino-Mazzucchelli,
  ``Conformal supergravity in three dimensions: Off-shell actions,''
JHEP {\bf 1310}, 073 (2013)
[arXiv:1306.1205 [hep-th]].


\bibitem{BKNT-M1}
 D.~Butter, S.~M.~Kuzenko, J.~Novak and G.~Tartaglino-Mazzucchelli,
  ``Conformal supergravity in three dimensions: New off-shell formulation,''
JHEP {\bf 1309}, 072 (2013)
[arXiv:1305.3132 [hep-th]].

\bibitem{ButterN=1} 
  D.~Butter,
  ``N=1 Conformal superspace in four dimensions,''
  Annals Phys.\  {\bf 325}, 1026 (2010)
  [arXiv:0906.4399 [hep-th]].
  
\bibitem{ButterN=2} 
  D.~Butter,
  ``N=2 Conformal superspace in four dimensions,''
  JHEP {\bf 1110}, 030 (2011)
  [arXiv:1103.5914 [hep-th]].  

\bibitem{HIPT}
  P.~S.~Howe, J.~M.~Izquierdo, G.~Papadopoulos and P.~K.~Townsend,
  ``New supergravities with central charges and Killing spinors in 2+1 dimensions,''
  Nucl.\ Phys.\  B {\bf 467}, 183 (1996)
  [arXiv:hep-th/9505032].
 
\bibitem{KLT-M11} 
  S.~M.~Kuzenko, U.~Lindstr\"om and G.~Tartaglino-Mazzucchelli,
  ``Off-shell supergravity-matter couplings in three dimensions,''
  JHEP {\bf 1103}, 120 (2011)
  [arXiv:1101.4013 [hep-th]].

\bibitem{vanN85}
P.~van Nieuwenhuizen,
``D = 3 conformal supergravity and Chern-Simons terms,''
Phys.\ Rev.\  D {\bf 32}, 872 (1985).
  
\bibitem{RvanN86} 
  M.~Ro\v{c}ek and P.~van Nieuwenhuizen,
  ``N $\geq$ 2 supersymmetric Chern-Simons terms as d = 3 extended conformal supergravity,''
  Class.\ Quant.\ Grav.\  {\bf 3}, 43 (1986).

\bibitem{NT} 
  M.~Nishimura and Y.~Tanii,
  ``N=6 conformal supergravity in three dimensions,''
  arXiv:1308.3960 [hep-th].


\bibitem{GreitzHowe} 
  J.~Greitz and P.~S.~Howe,
  ``Maximal supergravity in three dimensions: supergeometry and differential forms,''
  JHEP {\bf 1107}, 071 (2011)
  [arXiv:1103.2730 [hep-th]].
  
\bibitem{GGHN} 
U.~Gran, J.~Greitz, P.~Howe and B.~E.~W.~Nilsson,
``Topologically gauged superconformal Chern-Simons matter theories,''
JHEP {\bf 1212}, 046 (2012)
  [arXiv:1204.2521 [hep-th]].

\bibitem{NT12} 
  M.~Nishimura and Y.~Tanii,
  ``Coupling of the BLG theory to a conformal supergravity background,''
  JHEP {\bf 1301}, 120 (2013)
  [arXiv:1206.5388 [hep-th]].

\bibitem{LR89}
  U.~Lindstr\"om and M.~Ro\v{c}ek,
  ``Superconformal gravity in three dimensions as a gauge theory,''
  Phys.\ Rev.\ Lett.\  {\bf 62}, 2905 (1989).

\bibitem{NG}
  H.~Nishino and S.~J.~Gates Jr.,
  ``Chern-Simons theories with supersymmetries in three dimensions,''
  Int.\ J.\ Mod.\ Phys.\  A {\bf 8}, 3371 (1993).

\bibitem{CN} 
  X.~Chu and B.~E.~W.~Nilsson,
  ``Three-dimensional topologically gauged N=6 ABJM type theories,''
  JHEP {\bf 1006}, 057 (2010)
  [arXiv:0906.1655 [hep-th]].


\bibitem{KT-M12} 
 S.~M.~Kuzenko and G.~Tartaglino-Mazzucchelli,
 ``Conformal supergravities as Chern-Simons theories revisited,''
JHEP {\bf 1303}, 113 (2013)
  [arXiv:1212.6852 [hep-th]].  

\bibitem{ZP}
  B.~M.~Zupnik and D.~G.~Pak,
  ``Superfield formulation of the simplest three-dimensional gauge theories and
  conformal supergravities,''  Theor.\ Math.\ Phys.\  {\bf 77} (1988) 1070
  [Teor.\ Mat.\ Fiz.\  {\bf 77} (1988) 97].
 
\bibitem{Kuzenko12} 
S.~M.~Kuzenko,
``Prepotentials for N=2 conformal supergravity in three dimensions,''
JHEP {\bf 1212}, 021 (2012)  [arXiv:1209.3894 [hep-th]].

\bibitem{Aharony:2008ug} 
  O.~Aharony, O.~Bergman, D.~L.~Jafferis and J.~Maldacena,
  ``N=6 superconformal Chern-Simons-matter theories, M2-branes and their gravity duals,''
  JHEP {\bf 0810}, 091 (2008)
  [arXiv:0806.1218 [hep-th]].

\end{thebibliography}
\end{document}